\documentclass[cernpreprint,atlasdraft=true,texlive=2016,english,texmf]{atlasdoc}
 
\usepackage{atlaspackage}
\usepackage{atlasbiblatex}
 
\usepackage{atlasphysics}
 
\graphicspath{{logos/}{figures/}}
 
\usepackage{ANA-HDBS-2018-11-PAPER-defs}

\AtlasTitle{Search for resonances decaying into a weak vector boson and a Higgs boson
in the fully hadronic final state produced in proton--proton collisions at
$\sqrt{s} = 13$~TeV with the ATLAS detector}
 
\AtlasAbstract{
A search for heavy resonances decaying into a $W$ or $Z$ boson and a Higgs boson
produced in proton--proton collisions at the Large Hadron Collider at $\sqrt{s} = 13$~TeV is presented.
The analysis utilizes the dominant $W \to q \bar{q}^\prime$ or
$Z \to q \bar{q}$ and $H \to b \bar{b}$ decays
with substructure techniques applied to large-radius jets.
A sample corresponding to an integrated luminosity of 139~fb$^{-1}$ collected with the
ATLAS detector is analyzed and
no significant excess of data is observed over the background prediction.
The results are interpreted in the context of the heavy vector triplet model with
spin-1 $W^\prime$ and $Z^\prime$ bosons. Upper limits on the cross section are set for
resonances with mass between 1.5 and 5.0~TeV, ranging from \maxwhxs\ to \minwhxs~fb for
$W^\prime \to WH$ and from \maxzhxs\ to \minzhxs~fb for $Z^\prime \to ZH$ 
at the 95\% confidence level.
}
 
\author{The ATLAS Collaboration}
 
\AtlasRefCode{HDBS-2018-11}
 
\PreprintIdNumber{CERN-EP-2020-073}
 
\AtlasJournalRef{Phys. Rev. D 102 (2020) 112008}
 
\AtlasDOI{10.1103/PhysRevD.102.112008}
 
\AtlasJournal{Phys. Rev. D}

\hypersetup{pdftitle={ATLAS document},pdfauthor={The ATLAS Collaboration}}
 
\begin{document}
 
\maketitle

\section{Introduction}
% The next lines are included from the .//sections/sec-introduction.tex input file
The search for physics beyond the Standard Model (SM) is a major focus of the physics
program at the Large Hadron Collider (LHC).
Since its discovery~\cite{HIGG-2012-27, CMS-HIG-12-028}, the Higgs boson has become
a tool in this search. In particular, one may expect new heavy resonances to couple
to Higgs bosons and weak vector bosons ($V = W$ or $Z$).
Such resonances are expected to occur in a number of theories beyond the Standard Model.
Theories that aim to solve the naturalness problem predict the existence of vector
resonances as expected in composite Higgs models~\cite{Dugan:1984hq, Agashe:2004rs},
Little Higgs models~\cite{Schmaltz:2005ky},
or models with extra dimensions~\cite{Agashe:2007ki, Agashe:2008jb}.
Theories with extended Higgs sectors predict scalar resonances as in two-Higgs-doublet
models~\cite{Branco:2011iw}.
 
In this article, a search for $WH$ and $ZH$ resonances produced in proton--proton ($pp$)
collisions at $\rts = 13$~\TeV\ is reported with a sample corresponding to an integrated
luminosity of 139~\ifb\ collected with the ATLAS detector during Run 2 of the LHC in 2015--2018.
The search is designed for resonances with a mass of at least 1.5~\TeV\ and with both the $V$
and $H$ bosons decaying hadronically in the modes $V \to q\bar{q}^{(\prime)}$ and $H \to b\bar{b}$, as shown in Figure~\ref{fig:feyn}.
In this regime, the $V$ and $H$ bosons are produced with high transverse momentum (\pt),
resulting in each boson being reconstructed as a single large-radius hadronic jet, and the invariant mass of this dijet system provides the final discriminating variable. Jet substructure
techniques and $b$-tagging are then used to discriminate those jets from background jets
originating from multijet, $V$+jets, and \ttbar\ events~--~with QCD multijet events representing
at least 85\% of the total background. Due to difficulties in modeling
the background from simulation, all background estimates are derived from the data.
 
\begin{figure}[htb]
\begin{center}
\includegraphics[angle=-90,width=0.45\textwidth]{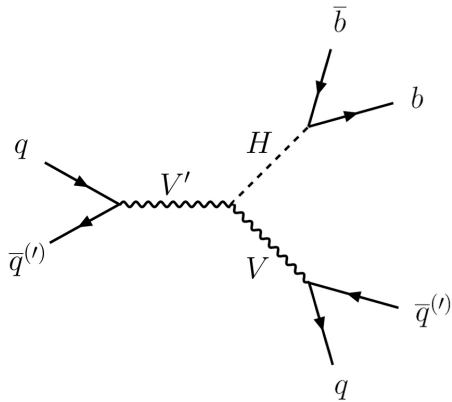}
\end{center}
\vspace{-0.6cm}
\caption{Feynman diagram for the production of a $V^\prime$ resonance with decay into a $VH$ pair.}
\label{fig:feyn}
\end{figure}
 
The results of the search are interpreted in the context of
the heavy vector triplet (HVT) model~\cite{Pappadopulo:2014qza},
which is a simplified model providing a broad
phenomenological framework for heavy resonances coupling to SM fermions and bosons.
In this model, $W^\prime$ and $Z^\prime$ vector bosons interact with quarks and the Higgs
field with coupling strength of $g_q$ and $g_H$, respectively.\footnote{Further details about the use of the HVT model in ATLAS analyses can be found in Ref.~\cite{EXOT-2017-31}.}
Coupling to the Higgs field
gives rise to interactions with longitudinally polarized $W$ and $Z$ bosons.
Two scenarios are considered as benchmarks for interpretation in this article.
Model A corresponds to the choice $g_q = -0.55$ and $g_H = -0.56$,
which reproduces the phenomenology of weakly coupled models based on
an extended gauge symmetry~\cite{Barger:1980ix}.
Model B corresponds to $g_q = 0.14$ and $g_H = -2.9$, which implements a strongly coupled
scenario as in composite Higgs models.
 
Previous searches for $VH$ resonances have been carried out at $\rts = 13$~\TeV\
in the semileptonic final state
($\nu\nu bb$, $\ell\nu bb$, and $\ell\ell bb$)~\cite{EXOT-2016-10, CMS-B2G-17-004, CMS-B2G-17-006}
and fully hadronic final state ($qq bb$)~\cite{EXOT-2016-12, CMS-B2G-17-002}.
The ATLAS and CMS collaborations report similar lower limits on the masses of
$W^\prime$ and $Z^\prime$ bosons in these two sets of final states with about 36~\ifb\ of integrated luminosity collected in 2015--2016.
The strongest lower limit on the $W^\prime$ mass is set by CMS in the $\ell\nu bb$
channel~\cite{CMS-B2G-17-004} with a value of 2.9~\TeV\ at the 95\% confidence level (CL)
in the context of HVT model B.
For $Z^\prime$ bosons, the strongest lower limit on the mass is set by ATLAS in the
combination of the $\nu\nu bb$ and $\ell\ell bb$ channels~\cite{EXOT-2016-10} with a value of 2.83~\TeV\ at the 95\% CL in HVT model B. 
 
The results presented in this article differ from those previously published by ATLAS in the
$qq bb$ channel~\cite{EXOT-2016-12} thanks to the following improvements.
The integrated luminosity has increased by a factor of nearly four,
an improved clustering algorithm combining measurements from the calorimeter and
tracking systems is used to reconstruct $V$- and $H$-candidate jets,
the $b$-tagging procedure used to identify $H$-candidate jets is performed on track-jets
with a \pt-dependent radius that allows double $b$-tagging of $H$-candidate jets up
to considerably higher \pt\ values,
and the Higgs-candidate selection has been reoptimized with increased sensitivity.

% End of text imported from the .//sections/sec-introduction.tex input file

\section{ATLAS detector}
% The next lines are included from the .//sections/sec-detector.tex input file
\newcommand{\AtlasCoordFootnote}{
ATLAS uses a right-handed coordinate system with its origin at the nominal interaction point (IP)
in the center of the detector and the $z$-axis along the beam pipe.
The $x$-axis points from the IP to the center of the LHC ring,
and the $y$-axis points upwards.
Cylindrical coordinates $(r,\phi)$ are used in the transverse plane,
$\phi$ being the azimuthal angle around the $z$-axis.
The pseudorapidity is defined in terms of the polar angle $\theta$ as $\eta = -\ln \tan(\theta/2)$.
The rapidity is defined relative to the beam axis as $y = (1/2) \ln [(E+p_z)/(E-p_z)]$.
Angular distance is measured in units of $\Delta R \equiv \sqrt{(\Delta\eta)^{2} + (\Delta\phi)^{2}}$.}
 
The ATLAS experiment~\cite{PERF-2007-01} at the LHC is a multipurpose particle detector
with a forward--backward symmetric cylindrical geometry and a near $4\pi$ coverage in
solid angle.\footnote{\AtlasCoordFootnote}
It consists of an inner detector (ID) for tracking surrounded by a thin superconducting solenoid
providing a \SI{2}{\tesla} axial magnetic field, electromagnetic (EM) and hadronic calorimeters, and a muon spectrometer (MS).
The inner detector covers the pseudorapidity range $|\eta| < 2.5$.
It consists of silicon pixel, silicon microstrip, and transition radiation tracking detectors.
An additional innermost pixel layer~\cite{ATLAS-TDR-2010-19,PIX-2018-001} inserted at a radius of 3.3~cm has been used since 2015.
Liquid-argon (LAr) sampling calorimeters provide EM energy measurements
with high granularity.
A hadronic scintillator-tile calorimeter covers the central pseudorapidity range ($|\eta| < 1.7$).
The endcap and forward regions are instrumented with LAr calorimeters
for both the EM and hadronic energy measurements up to $|\eta| = 4.9$.
The muon spectrometer surrounds the calorimeters and features
three large air-core toroidal superconducting magnet systems with eight coils each.
The field integral of the toroids ranges between \num{2.0} and \SI{6.0}~Tm
across most of the detector. The muon spectrometer includes a system of precision tracking
chambers and fast detectors for triggering.
A two-level trigger system~\cite{TRIG-2016-01} is used to select events.
The first-level trigger is implemented in hardware and uses a subset of the detector information
to reduce the accepted rate to at most \SI{100}{\kilo\hertz}.
This is followed by a software-based trigger level that reduces the accepted event rate
to \SI{1}{\kilo\hertz} on average.
% End of text imported from the .//sections/sec-detector.tex input file

\section{Data and Monte Carlo simulation}
% The next lines are included from the .//sections/sec-data-MC.tex input file
The data sample for the analysis was collected by the ATLAS detector
with high-\pt\ single-jet triggers utilizing the \antikt\ clustering algorithm~\cite{Cacciari:2008gp}
with a radius $R = 1.0$. The lowest unprescaled triggers were used with the following
\pt\ thresholds: 360~\GeV\ in 2015, 420~\GeV\ in 2016, and 460~\GeV\ in 2017--2018. After requiring that the data were recorded with stable beam conditions and
satisfied detector and data quality requirements, the integrated luminosity was measured to be 139~\ifb\
using the methodology from Ref.~\cite{ATLAS-CONF-2019-021}.
 
The analysis relies on Monte Carlo (MC) samples to model signal events. Background
MC samples are used only to optimize the signal event selection and to validate the data-driven
background estimation method (Section~\ref{sec:background}).
 
Signal $W^\prime$ and $Z^\prime$ events for HVT model A were produced at leading-order (LO)
precision in the strong coupling constant ($\alpha_\textrm{s}$) with the
\MGMCatNLO 2.2.2~\cite{Alwall:2011uj} event generator using the NNPDF23LO parton
distribution function (PDF) set~\cite{Ball:2012cx}. Separate generation of signal events for HVT model B is not required as both models A and B give rise to dijet mass peaks with a width that is dominated by the experimental resolution. The events were interfaced with
\PYTHIAV{8.186}~\cite{Sjostrand:2007gs} for parton showering, hadronization, and
the underlying event, and relied on the A14 set of tuned parameters~\cite{ATL-PHYS-PUB-2014-021}.
Higgs boson decays to heavy-flavor final states $H \to b\bar{b}$ and $H \to c\bar{c}$
were included, corresponding to branching fractions of 58.2\% and 2.9\%, respectively~\cite{deFlorian:2016spz}.
 
Multijet events were produced with the \PYTHIAV{8.186} event generator, the
NNPDF23LO PDF set, and the A14 tune.
Samples of events with top-quark pairs were produced at next-to-leading order (NLO)
with \POWHEGBOX~\cite{Frixione:2007nw} and the NNPDF30 NLO PDF set,
interfaced with \PYTHIAV{8.183} and the A14 tune.
The $h_\textrm{damp}$ parameter was set to 1.5 times the top-quark mass~\cite{ATL-PHYS-PUB-2014-005}.
Samples of $W$+jets and $Z$+jets events were produced with \SHERPA 2.1.1~\cite{Gleisberg:2008ta}
and the CT10 PDF set~\cite{ct10} for up to two partons at NLO and
up to four partons at LO.
The cross sections used to normalize the multijet and the $V$+jets MC samples were computed with \PYTHIA and \SHERPA, respectively, and the top-quark pair cross section was taken to be $832^{+46}_{-52}$~pb for a top-quark
mass of 172.5~\GeV. This value was calculated at next-to-next-to-leading order in $\alpha_\textrm{s}$,
including the resummation of next-to-next-to-leading logarithmic soft gluon terms, with
Top++2.0~\cite{Cacciari:2011hy,Beneke:2011mq,Baernreuther:2012ws,Czakon:2012zr,Czakon:2012pz,Czakon:2013goa,Czakon:2011xx}. Other SM backgrounds originating from diboson production ($VV$) and weak vector-boson production in association with a Higgs boson ($VH$) are negligible and not considered.
 
For all MC samples, except those produced with \SHERPA,
$b$-hadron and $c$-hadron decays were handled by {\textsc{EvtGen}} 1.2.0~\cite{Lange:2001uf}.
Inelastic $pp$ events generated using \PYTHIAV{8.186} with the A3 tune~\cite{ATL-PHYS-PUB-2016-017}
and the NNPDF23LO PDF set
were added to the hard-scattering interaction in such a way as to reproduce the effects of additional
$pp$ interactions (pileup) in each bunch crossing during data collection.
The detector response was simulated with \GEANT~4~\cite{geant, SOFT-2010-01},
and the events were processed with the same reconstruction software as for the data.
Energy/momentum scale and efficiency corrections are applied to the results of the simulation to account
for small differences between the simulation and the performance measured directly from the
data.

% End of text imported from the .//sections/sec-data-MC.tex input file

\section{Event reconstruction and selection}
% The next lines are included from the .//sections/sec-reco-selec.tex input file
 
The analysis relies on the reconstruction of charged particles with $\pt > 500$~\MeV\
in the inner detector to reconstruct $pp$ collision vertices for each crossing of the
LHC beams. The primary vertex is chosen to be the vertex with the largest $\sum \pt^2$
for the tracks associated with the vertex.
 
Jets are built from a combination of tracks and calibrated clusters of energy deposits in calorimeter cells~\cite{PERF-2011-03}, with the \antikt\ clustering algorithm using a radius parameter $R = 1.0$ as implemented in \textsc{FastJet}~\cite{Cacciari:2011ma}. The tracks are selected with the same requirements as in Ref.~\cite{ATLAS-CONF-2016-039}, except for the minimum \pt\ value, which has been increased to 500~\MeV.
By combining calorimeter and tracking information, one benefits from both the better
energy resolution of the calorimeter at high energy and the superior angular resolution for the tracks.
This combination becomes highly beneficial at large jet \pt\ due to the small number
of clusters produced and the limited angular resolution of the calorimeter.
The resulting jets are referred to as Track-CaloCluster (TCC) jets. A detailed
description of the algorithm can be found in Ref.~\cite{ATL-PHYS-PUB-2017-015}
and its application to a search for high-mass diboson resonances is described in Ref.~\cite{HDBS-2018-31}.
A trimming algorithm~\cite{Krohn2010} is applied to minimize the impact of pileup.
In this algorithm, the constituents of each jet are reclustered
with the \kt\ algorithm~\cite{Ellis:1993tq} into smaller $R = 0.2$ subjets. Trimmed large-$R$ jets are made up of constituents of those subjets with $\pt^{\text{subjet}} / \,\pt^{\text{jet}} > 0.05$, where $\pt^{\text{subjet}}$
and $\pt^{\text{jet}}$ are the transverse momenta of the subjet and original
untrimmed jet, respectively.
The energy and mass calibration of TCC jets is based on the simulation as described
in Ref.~\cite{ATL-PHYS-PUB-2015-033}.
As a result of the improved angular resolution of the energy distribution within the jet,
the discrimination between signal $W$ or $Z$ jets and background QCD jets is noticeably
improved. In addition to their masses, a powerful variable to discriminate between those jets is
\dtwo, defined as the ratio of three-point to two-point energy correlation functions
that are based on the energies of the jet constituents and their pairwise angular
separation~\cite{Larkoski:2013eya,Larkoski:2014gra}. The \dtwo~variable exploits the two-body structure of the $V \to q\bar{q}^{(\prime)}$ decays, absent from typical QCD jets.
Another variable that provides discrimination between signal and background jets is the
number of tracks (\ntrk) matched to the jets by ghost association~\cite{Cacciari:2008gn}. This quantity is significantly higher for gluon-induced jets that are a component of the background than for quark-induced jets in signal events, due to the distinct energy scales involved and the different color factors for gluons and quarks.
 
To identify Higgs-boson jets, a separate collection of jets is built from tracks with the \antikt\ algorithm using a \pt-dependent radius
$R = \rho / \pt$~\cite{Krohn:2009zg}, where the parameter $\rho$ is set to 30~\GeV\
and the radius is constrained to remain in the range between 0.02 and 0.4~\cite{ATL-PHYS-PUB-2017-010}.
The track-jets are assigned to specific large-$R$ jets by ghost association
with the original untrimmed large-$R$ jets.
The main advantage of using such variable-radius track-jets is that one can resolve
the track-jets from $H \to b\bar{b}$ decays at high \pt\ and
retain the ability to double $b$-tag the large-$R$ Higgs-candidate jets.
 
Track-jets are tagged as likely to contain $b$-hadrons if they satisfy the selection criteria
of the MV2c10 algorithm~\cite{PERF-2012-04,ATL-PHYS-PUB-2016-012} that takes advantage of
the relatively long lifetime and large mass of $b$-hadrons.
A working point corresponding to a $b$-tagging efficiency of 77\% for true $b$-jets is used.
For this efficiency value, rejection factors of 5 and 110 are obtained against $c$-quark and
light-quark jets, with the efficiency and rejection factors determined in \ttbar\ MC simulation~\cite{FTAG-2018-01}.
 
Electrons are reconstructed by matching ID tracks to energy clusters in the EM calorimeter.
The identification of electrons relies on a likelihood discriminant that takes the characteristic
shape of electromagnetic showers into account~\cite{EGAM-2018-01}.
Electrons are required to have $\pt > 7$~\GeV\ and $|\eta| < 2.47$, and satisfy the \textit{loose}
identification criteria~\cite{EGAM-2018-01}.
The associated tracks must have a transverse impact parameter significance
$|d_0|/\sigma_{d_0} < 5$ relative to the beam axis and a longitudinal impact parameter $|z_{0} \sin{\theta}|< 0.5$~mm relative to the primary vertex.
Muons are reconstructed and identified by matching ID and MS tracks, and performing a global
fit with all ID and MS measurements, taking the energy loss in the calorimeter into account~\cite{PERF-2015-10}.
Muons are required to have $\pt > 7$~\GeV\ and $|\eta| < 2.5$, and satisfy the \textit{loose} identification
criteria~\cite{PERF-2015-10}.
The following track requirements are applied: $|d_0|/\sigma_{d_0} < 3$ and
$|z_0 \sin\theta| < 0.5$~mm.
Both the electrons and muons are required to satisfy loose isolation criteria~\cite{EGAM-2018-01,PERF-2015-10}
based on the total transverse momentum of tracks surrounding the leptons within a cone of radius
$\Delta R = \min(10\,/\,\pt^\ell[\textrm{\GeV}], \Delta R_\textrm{max})$, where $\pt^\ell$ is the lepton \pt, and
$\Delta R_\textrm{max} = 0.2$ for electrons and 0.3 for muons.
The isolation criteria have an efficiency of 99\% with negligible dependence on the lepton \pt\ value.
 
Events must satisfy the trigger requirements and contain a primary vertex.
In addition, non-collision backgrounds originating from calorimeter noise, beam halo
interactions or cosmic rays are suppressed by rejecting events that contain
any $R = 0.4$ \antikt\ calorimeter jet failing to satisfy a set of quality criteria. These are based on the LAr pulse shape, the energy profile of the jet in different parts of the calorimeter, and track variables~\cite{ATLAS-CONF-2015-029}.
Events with one or more charged leptons (electrons or muons) are also rejected to retain orthogonality
with other $VH$ search channels.
 
The signal topology requires the presence of two large-$R$ jets with $|\eta| < 2.0$ and
$\pt > 200$~\GeV. The leading (highest \pt) jet must satisfy $\pt > 500$~\GeV. The invariant mass
of the dijet system consisting of the two highest-\pt\ jets in the event (\mjj) is required to be
larger than 1.3~\TeV. These kinematic requirements guarantee that the trigger is fully efficient.
To suppress $t$-channel multijet production, the difference between the rapidities of the two
leading jets must satisfy $|\Delta y| < 1.6$. Only the two leading-\pt\ jets are retained
for further consideration.
 
As an initial step, the jet with the larger mass is taken to be the $H$-boson candidate while
the other jet is taken to be the $V$-boson candidate.
Discrimination between these $V$-boson or $H$-boson candidates and background jets
relies on several properties of the large-$R$ jets: mass, \dtwo, and \ntrk.
An optimization procedure is applied to adjust the selection criteria involving those variables to
maximize the significance of the resonance signal under study. In the case of $V$-boson jets,
the selection is based on the three discriminating variables as developed
in the search for heavy diboson resonances in the fully
hadronic channel~\cite{HDBS-2018-31}, with the exception that the \ntrk\ selection is loosened slightly.
In the case of the $H$-boson candidates, the selection criteria are optimized with regard to the jet mass and \ntrk. The successful $H$-boson candidate has at least one associated track-jet and can be classified as either 1-tag or 2-tag, depending on the number of
track-jets satisfying the $b$-tagging requirements. Only the two highest-\pt\ associated track-jets are considered for $b$-tagging. The variable \dtwo\ provides little additional discrimination and is thus dropped.
 
The \pt-dependent jet mass windows for the $W$($Z$)-boson candidates that result from the optimization procedure vary from 80--100~(85--110)~\GeV\ for jets with \pt\ around 500~\GeV\ to 55--130~(65--135)~\GeV\ for jets with \pt\ around 3000~\GeV. The upper bounds on \dtwo\ for the $V$-boson candidates vary approximately from 1.0 to 1.5, for the same \pt\ values. The upper cut on the \ntrk\ variable varies between 25 (26) and 31 (29) for $W$($Z$)-boson candidates, with looser requirements for higher-\pt\ jets. Given the experimental jet mass resolution, no exclusive selection of $W$ or $Z$ bosons is performed and the $WH$ and $ZH$ final states are searched for independently.
For $H$-boson candidates, the mass windows applied to events in the 1-tag (2-tag) category vary from 80--135 (95--150)~\GeV\ to 105--155 (100--170)~\GeV\ for jets with \pt\ of 500 and 3000~\GeV, respectively. The $H$-boson candidates classified as 1(2)-tag are required to have an associated \ntrk\ value below 32 (35) to 44 (55), looser at higher \pt.
 
The event selection defining the signal region (SR) is summarized in Table~\ref{tab:selection}.
The resulting signal acceptance times efficiency (\acceff) for events in each category is shown in Figure~\ref{fig:eff}. In the 1-tag category, \acceff\ rises from approximately 3\% to 10\% for $WH$ resonance masses increasing from 1.5 to 5.0~\TeV.
In the 2-tag category, \acceff\ remains essentially constant at 4\% for $WH$ resonances. The different trends for the 1-tag and 2-tag selections as a function of $m(V')$ are the result of a combination of effects including the \pt\ dependence of both $V$- and $H$-tagging as well as the signal to background ratio. The \acceff\ values are about 0.5\% lower for $ZH$ resonances due to the smaller mass separation between $Z$ and $H$ bosons. The experimental mass resolution for resonance masses of 2 (4)~\TeV\ is 3.5\% (2.6\%).
 
\begin{figure}[tb]
\begin{center}
\includegraphics[width=0.55\textwidth]{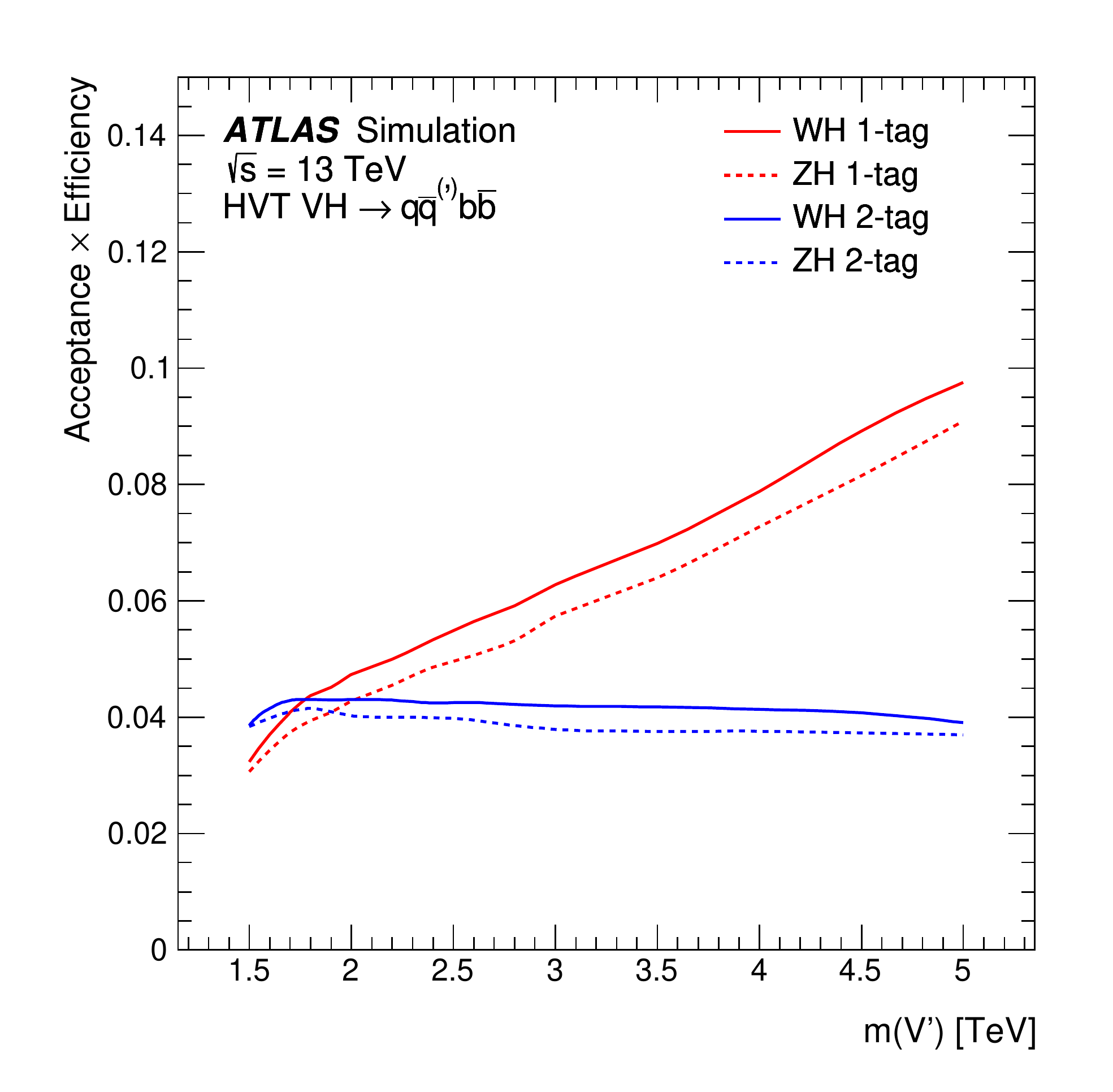}
\end{center}
\vspace{-0.6cm}
\caption{Signal acceptance times efficiency as a function of the resonance mass, for events in the $WH$ (solid lines) and $ZH$ (dashed lines) signal regions, in the 1-tag and 2-tag categories, with respect to the total number of generated events in each sample. The HVT MC samples include only $V^\prime \rightarrow VH$ decays with $V \to q\bar{q}^{(\prime)}$ and $H \to b\bar{b}$ or $c\bar{c}$. }
\label{fig:eff}
\end{figure}
 
\begin{table}[htp]
\caption{Event selection requirements and definition of the different regions used in the
analysis. Events in the signal, control, and validation regions must satisfy the preselection requirements.
}
\begin{center}
\begin{tabular}{l  l}
\hline\hline
Preselection       & Veto non-$qqqq$ channels: \\
& ~~~~~~ No $e$ ($\mu$) with $\pt > 7$~\GeV{} and $|\eta| < 2.47$ ($2.5$) \\
& Event kinematics: \\
& ~~~~~~ $\geq 2$ large-$R$ jets with $\pt > 200$~\GeV{} and $|\eta| < 2.0$ \\
& ~~~~~~ leading large-$R$ jet with $\pt > 500$~\GeV{} \\
& ~~~~~~ leading and subleading large-$R$ jets with $\mjj > 1.3$~\TeV{} \\
& ~~~~~~ leading and subleading large-$R$ jets with $|\Delta y| < 1.6$ \\
\midrule
$V/H$ assignment    & $V$-boson ($H$-boson) candidate is large-$R$ jet with lower (higher) mass \\
\midrule
Signal region      & $V$ and $H$ bosons: \\
(SRWH / SRZH)      & ~~~~~~ $W$-boson candidate within $W$ jet mass, \dtwo, and \ntrk\ windows \\
& ~~~~~~ $Z$-boson candidate within $Z$ jet mass, \dtwo, and \ntrk\ windows \\
& ~~~~~~ $H$-boson candidate within $H$ jet mass, \ntrk\ windows, with 1 or 2 $b$-tagged track-jets \\
\midrule
Control region      & Fail both SRWH and SRZH \\
(CR)                & Pass $H$-boson candidate \ntrk \\
& $(m_V < 65~\GeV{} \,\&\ m_H < 70~\GeV{})$ or $(m_V > 110~\GeV{} \,\&\ m_H > 150~\GeV{})$ or \\
& $(m_V < 65~\GeV{} \,\&\ m_H > 150~\GeV{})$ \\
\midrule
Validation region   & Fail both SRWH and SRZH \\
(VR1A)              & Pass $V$-boson candidate \ntrk \\
& Pass $H$-boson candidate \ntrk \\
& $65 < m_V < 110~\GeV{} \,\&\ m_H > 150~\GeV{}$ \\
\midrule
Validation region   & Fail both SRWH and SRZH \\
(VR1B)              & Fail $V$-boson candidate \ntrk \\
& Pass $H$-boson candidate \ntrk \\
& $65 < m_V < 110~\GeV{} \,\&\ m_H > 150~\GeV{}$ \\
\midrule
Validation region   & Fail both SRWH and SRZH \\
(VR2A)              & Pass $V$-boson candidate \ntrk \\
& Pass $H$-boson candidate \ntrk \\
& $m_V < 65~\GeV{} \,\&\ 70 < m_H < 150~\GeV{}$ \\
\midrule
Validation region   & Fail both SRWH and SRZH \\
(VR2B)              & Fail $V$-boson candidate \ntrk \\
& Pass $H$-boson candidate \ntrk \\
& $m_V < 65~\GeV{} \,\&\ 70 < m_H < 150~\GeV{}$ \\
\hline\hline
\end{tabular}
\label{tab:selection}
\end{center}
\end{table}
% End of text imported from the .//sections/sec-reco-selec.tex input file

\section{Background estimation and event yields}
\label{sec:background}
% The next lines are included from the .//sections/sec-background.tex input file
The dominant background in this search comes from multijet events, corresponding to at least 85\% of the total background in the signal regions, where the remaining events come from $t\bar{t}$ and $V$+jets processes.
 
The totality of the background is estimated via a data-driven method that provides template \mjj\ distributions for the $WH$ and $ZH$ final states in the 1-tag and 2-tag categories. These background \mjj\ templates are obtained in three steps: (i) background templates for the $WH$ and $ZH$ final states are extracted from data in the 0-tag category, where the $H$-boson candidate has zero associated $b$-tagged jets; (ii) yield and shape corrections are derived from a control region and applied to these templates; and (iii) a rebinning and smoothing of the resulting \mjj\ distributions is performed, to ensure robustness against statistical fluctuations. All steps are described in this section.
 
To define the 0-tag templates from which the background in the 1-tag and 2-tag categories is extracted, additional requirements are placed on 0-tag events, such that at least 1(2) variable-radius track-jet(s) is(are) associated with the $H$-boson jet, when estimating the background in the 1(2)-tag category. Events in the 0-tag category are not expected to directly describe either the yield or the \mjj\ shape of the background in the 1-tag and 2-tag categories without further corrections: the $b$-tagging requirements enhance heavy-flavor components in the background and introduce kinematics-dependent effects.
 
Therefore, a control region (CR) is used to estimate the yield and shape corrections to the \mjj\ distribution needed to extrapolate the 0-tag background events to the 1(2)-tag SRs. The CR has negligible contamination from signal. Validation regions (VR) with events that fail the SR selection are used to confirm the effectiveness of the background model and derive the associated systematic uncertainties. The definitions of the control and validation regions are shown in detail in Table~\ref{tab:selection} and illustrated in Figure~\ref{fig:regions}. According to simulation, multijet processes are responsible for 85\% to 99\% of the background composition across the different control, validation, and signal regions. This variation and its impact on the background estimate are taken into account by the uncertainties described in Section~\ref{sec:systematics}.
 
\begin{figure}[tb]
\begin{center}
\includegraphics[width=0.7\textwidth]{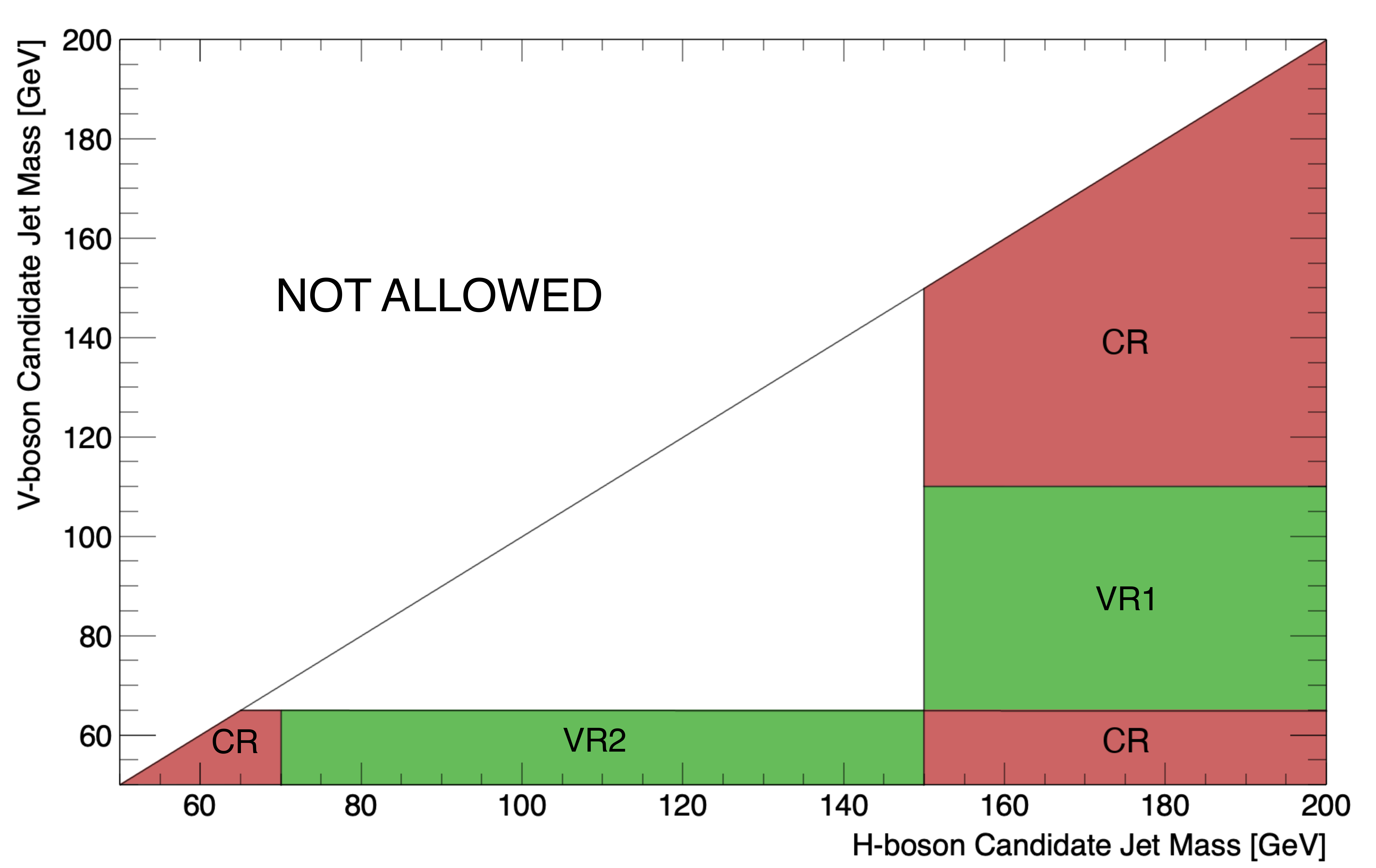}
\end{center}
\caption{Illustration of control and validation regions, defined by the masses of the $H$-boson and $V$-boson candidates. The regions VR1 and VR2 are further split into two regions each, according to the \ntrk\ requirement on the $V$-boson candidate. The same definitions are applied across the number of $b$-tags (0-, 1-, and 2-tag categories).}
\label{fig:regions}
\end{figure}
 
To define the background normalization in the 1(2)-tag category, a normalization correction $\mu^{1(2)\text{-tag}}_{\text{CR}}$ is extracted from the ratio of 1(2)-tag to 0-tag yields in the control region:
\begin{equation}\nonumber
\mu^{1(2)\text{-tag}}_{\text{CR}}=\frac{N^{1(2)\text{-tag}}_{\text{CR}}}{N^{0\text{-tag}}_{\text{CR}}},
\end{equation}
where $N^{0\text{-tag}}_{\text{CR}}$ and $N^{1(2)\text{-tag}}_{\text{CR}}$ are the numbers of events observed in the 0-tag and 1(2)-tag CR event categories. The values of $\mu^{1\text{-tag}}_{\text{CR}}$ and $\mu^{2\text{-tag}}_{\text{CR}}$ are determined to be $0.160\pm0.014$ and $0.0167\pm0.0028$, respectively, where the uncertainties are dominated by systematic effects discussed in Section~\ref{sec:systematics}. The difference in the corrections for 1-tag and 2-tag events can be understood based on studies of simulated multijet events. In 1-tag events, the two leading track-jets associated with Higgs candidate large-$R$ jets are dominated by one true $b$-jet accompanied by one true light-jet, whereas in 2-tag events they are dominated by two true $b$-jets. This indicates that different processes are at work in 1-tag and 2-tag events. The expected number of background 1(2)-tag events in the SR is calculated by applying the $\mu^{1(2)\text{-tag}}_{\text{CR}}$ correction factor to the number of 0-tag events that pass all SR requirements except for the $b$-tagging. 
 
To extract \mjj\ background templates from the 0-tag category, a multidimensional kinematic reweighting~\cite{hep-ml} is performed using the control region events. A boosted decision tree (BDT) is used to perform the reweighting, by predicting the event weights needed to bring the shapes of kinematic distributions in the 0-tag and 1(2)-tag categories into agreement. The training is performed in the CR data and performed separately for 1-tag and 2-tag events. Variables that are sensitive to the presence of $b$-jets associated with the $H$-boson candidate and to the resulting kinematic differences are used to train the BDTs: the four-momenta of the two leading variable-radius track-jets and their angular separation, the transverse momenta of the $H$- and $V$-boson jets, and the number of tracks and variable-radius track-jets associated with the $H$-boson jet. The BDTs are built with 100 trees, a maximum depth of three layers, and a minimum of 500 events per leaf, with a learning rate set to 0.1.
 
In order to quantify the effectiveness of the reweighting, a binary classifier was trained to differentiate between the reweighted and target \mjj\ distributions in the validation regions, and observed to classify them correctly at most 53\% of the time, consistent with random guessing. The observed distributions of kinematic variables, including \mjj, are found to be well described by the background model for 1-tag and 2-tag events in VR1A, VR1B, VR2A, and VR2B (defined in Table~\ref{tab:selection}).
 
The modeling of the \mjj\ distributions in the VR2B region is shown in Figure~\ref{fig:mVH_VR}, for 1-tag and 2-tag events. A residual disagreement between the data and the expected background after reweighting is accounted for by a systematic uncertainty, as discussed in Section~\ref{sec:systematics}.
 
\begin{figure}[t]
\begin{center}
\includegraphics[width=0.48\textwidth]{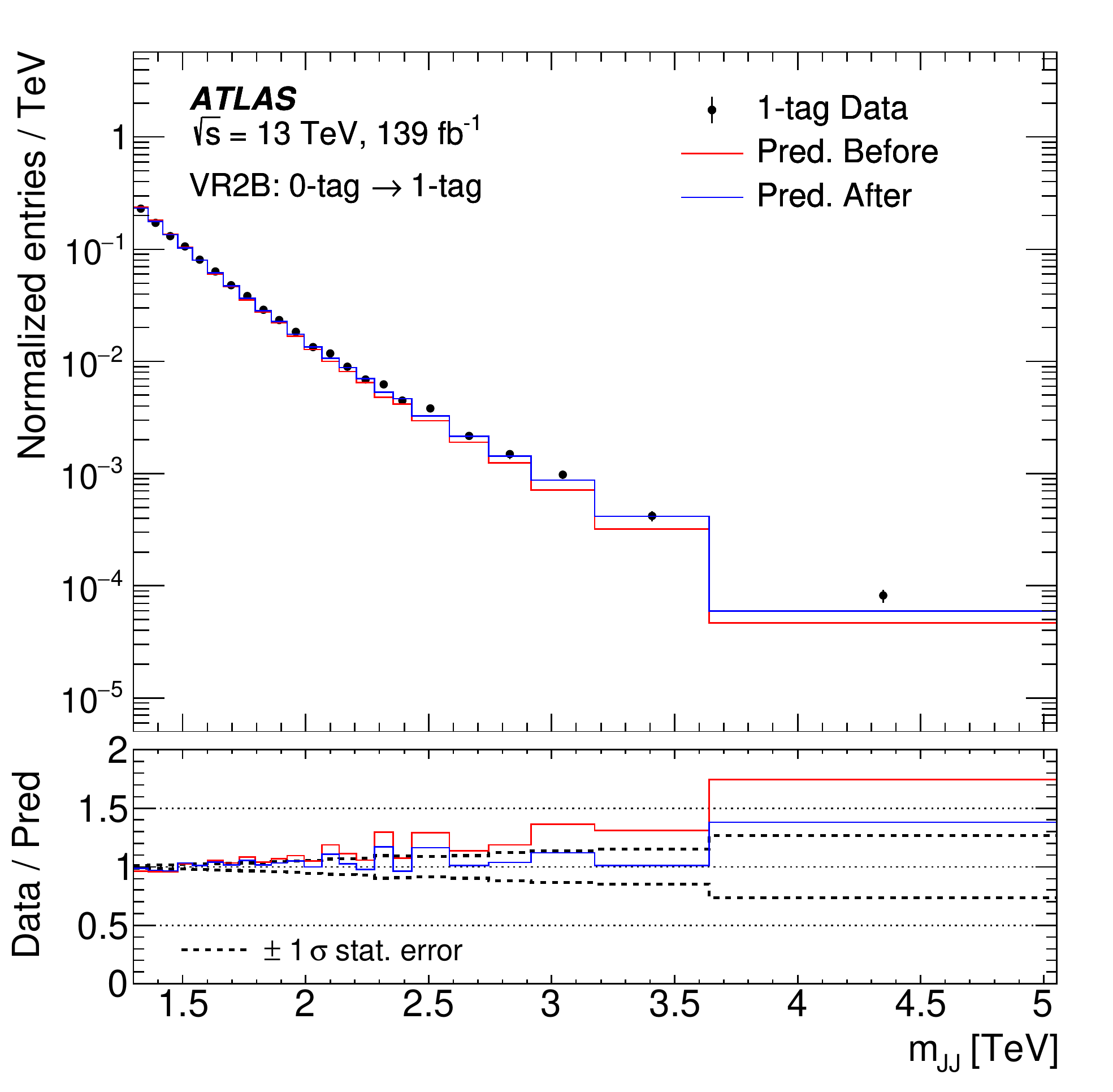}
\includegraphics[width=0.48\textwidth]{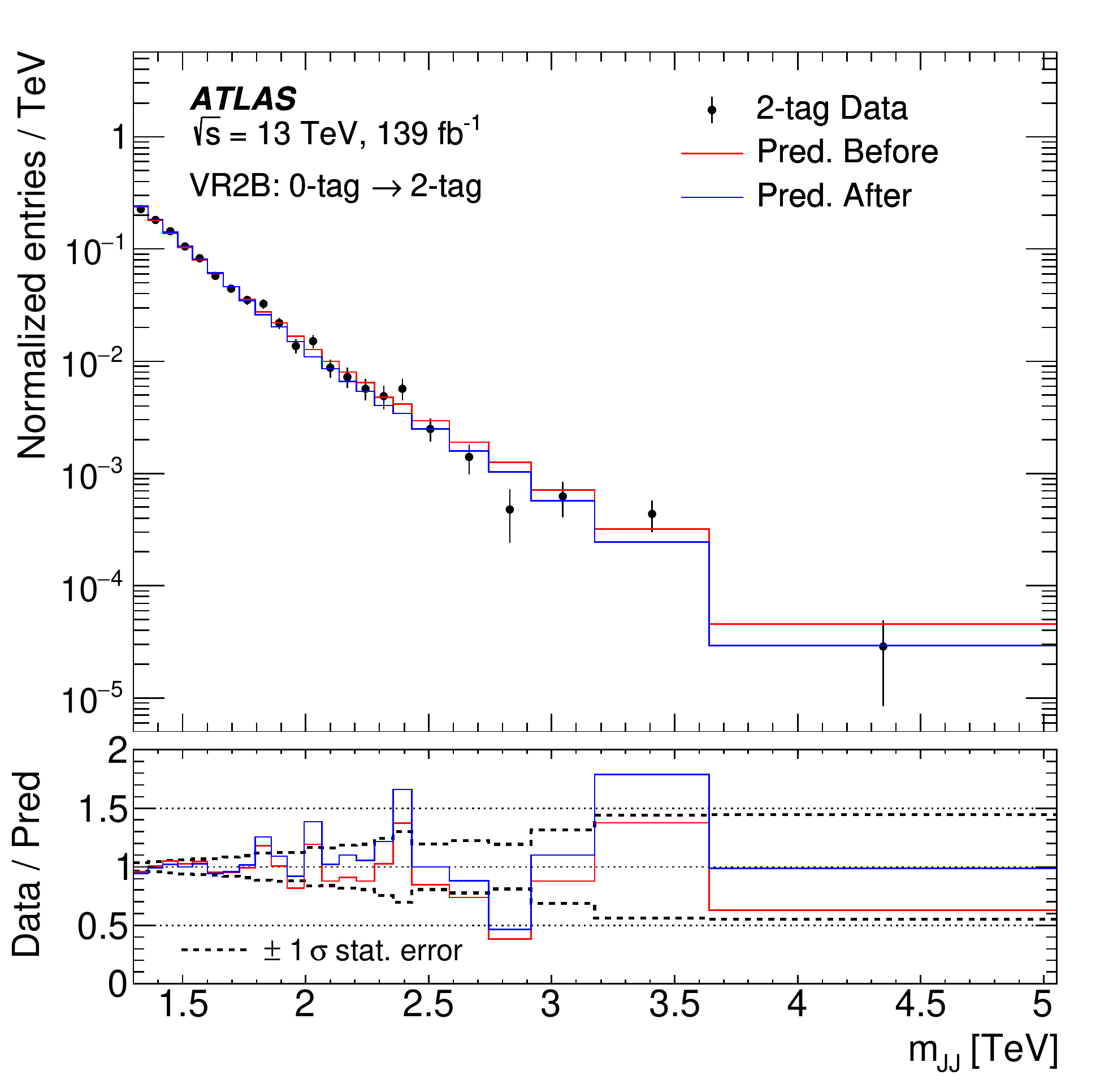}
\end{center}
\caption{Dijet mass distributions in the 1-tag (left) and 2-tag (right) VR2B regions compared with the predicted background extracted from the 0-tag events (histograms) before and after BDT reweighting. 
}
\label{fig:mVH_VR}
\end{figure}
 
The numbers of 1-tag and 2-tag events observed in the control and validation regions are shown in Table~\ref{tab:yields} and compared with the predicted background yields. 
 
\begin{table}[t]
\caption{Data and estimated background yields for 1-tag and 2-tag events in the control and validation regions. The uncertainties correspond to the combination of statistical and systematic components. By construction, the uncertainties cover the differences between the observed and expected yields in the validation regions.}
\begin{center}
\begin{tabular}{ccc}
\hline
\hline
1-tag & Data & Background prediction \\
\hline
Control region (CR)~ & $48668$ & $--$\\
Validation region (VR1A) &  ~~$4440$ &  $4290 \pm 380$\\
Validation region (VR1B) & $18361$ & $18000 \pm 1600$ \\
Validation region (VR2A) & $11844$ & $12600 \pm 1100$\\
Validation region (VR2B) & $29436$ & $31200 \pm 2700$\\
\hline
2-tag & Data & Background prediction \\
\hline
Control region (CR)~ & $4976$ & $--$ \\
Validation region (VR1A) & ~~$507$ &  $443 \pm 74$\\
Validation region (VR1B) & $1922$ & $1860 \pm 310$\\
Validation region (VR2A) & $1337$ & $1290 \pm 220$\\
Validation region (VR2B) & $3065$ & $3200 \pm 540$\\
\hline
\hline
\end{tabular}
\label{tab:yields}
\end{center}
\end{table}
 
After the normalization and reweighting corrections are applied to the events in the 0-tag category, the expected \mjj\ background distributions in 1-tag and 2-tag categories are produced with a variable bin width that reflects the experimental mass resolution. Those distributions are then fit using a functional form that captures the smoothly falling behavior of the background:
\begin{equation}\nonumber
f_{\text{Background}}(x)=\text{e}^{-p_{0}}(1-x)^{-p_{1}}x^{-p_{2}},
\end{equation}
where $x=\mjj/\sqrt{s}$ and $p_{0}$, $p_{1}$, and $p_{2}$ are the fit parameters.
The results of these fits provide the background estimates that are used in the statistical analysis (described in Section~\ref{sec:result}) for the different signal regions.
% End of text imported from the .//sections/sec-background.tex input file

\section{Systematic uncertainties}
\label{sec:systematics}
% The next lines are included from the .//sections/sec-systematics.tex input file
Systematic uncertainties arise from several different sources: the data-driven background estimate, the modeling of experimental uncertainties affecting the signal, and the impact of signal theory uncertainties. At a resonance mass of 2~\TeV, the background normalization and shape uncertainties dominate, while at a resonance mass of 4~\TeV, the large-radius jet and boson tagging uncertainties dominate due to the small background contribution at high \mjj.
 
The uncertainties affecting both the normalization and shape of the background predictions are determined from the validation regions. These uncertainties arise from limited sample sizes and from differences in the background composition in the various regions. The normalization uncertainty is taken to be the difference between the smallest and largest $\mu^{1(2)\text{-tag}}$ values obtained in any of the validations regions, resulting in a systematic error of 9\% (17\%) in the background estimate for the 1(2)-tag category.
The uncertainty in the shape of the background \mjj\ distribution is assessed from the ratio of data to prediction in the VR2B region, where the differences are the largest. This uncertainty is determined after smoothing the \mjj\ distributions for both the data and the background prediction with the same functional form as described in Section~\ref{sec:background}. It results in changes to the background yield of approximately 5\% at \mjj\ of 2~\TeV\ and up to 24\% at 4~\TeV.

An additional shape uncertainty is assigned to account for the choice of fitting function, assessed by fitting alternate empirical functions, amounting to a maximum uncertainty of 2\% (14\%) at an \mjj\ value of 2~(4)~\TeV.
 
Experimental uncertainties related to MC simulation are applicable only to signal samples, and are divided into two categories: $b$-tagging and large-$R$ jets. A set of $b$-tagging correction factors and corresponding uncertainties are applied as a function of \pt\ and $\eta$ of the variable-radius track-jets to match the efficiencies for tagging $b$-jets measured in data, determined with $t\bar{t}$ events~\cite{FTAG-2018-01}. The uncertainties in the correction factors are extrapolated for track-jets with $\pt$ larger than 400~\GeV. An additional extrapolation uncertainty is obtained by varying the inputs to the $b$-tagging algorithm according to their modeling uncertainties and by recomputing its efficiency in MC simulation~\cite{FTAG-2018-01}.
 
Uncertainties in the $\pt$ and mass scales of the large-$R$ jet are determined with the $R_\text{trk}$ method~\cite{JETM-2018-02} adapted to the jet collection used in this article, relying on independent measurements by the calorimeter and the inner detector, and are of the order of 5\% to 10\% each. These uncertainties lead to shifts in \mjj\ of the resonant signal peak as well as differences in the signal selection efficiency. The effects of resolution uncertainties on the $\pt$ and mass measurements are estimated by degrading the resolution in MC simulation according to a Gaussian smearing of width $0.02$ in $\sigma(\pt)/\pt$ and $0.20\times\sigma(m)/m$ in $\sigma(m)/m$~\cite{PERF-2012-02,ATLAS-CONF-2016-035}.
 
The $H$- and $V$-boson tagging techniques are assigned dedicated uncertainties to take into account the requirements on \dtwo~and \ntrk. An MC efficiency correction factor of $0.92\pm0.13$ for $V$-boson tagging was determined in the search for heavy diboson resonances in the fully hadronic channel~\cite{HDBS-2018-31}, by taking advantage of a control region in data that is enhanced in $V$+jets events. Given that the same method for $V$-tagging is used in this analysis, the same scale factor and uncertainty of 14\% is assigned to the signal normalization.\footnote{The only difference is in the \ntrk\ requirement, which is looser in this analysis. Studies of relative signal efficiencies for the different \ntrk\ requirements show that this approximation is well motivated.} An efficiency correction and uncertainty in the \ntrk\ requirement was also estimated in Ref.~\cite{EXOT-2016-19} for $V$-jets. This uncertainty is applied to $H$-tagging, with an additional component to cover topology differences, based on simulation studies, corresponding to a total 10\% uncertainty. In particular, these studies compare the large-radius jet mass distributions between the data and the simulation, and the impact of the \ntrk\ requirement on the data-to-MC agreement in the $V$-mass and $H$-mass regions.
 
Signal cross sections computed at leading order are used in the interpretation of the results. The impact of uncertainties in the PDF sets, initial- and final-state radiation, and multiparton interactions on the signal acceptance are included. Uncertainties related to the PDF sets are derived by applying the methodology outlined by the PDF4LHC group~\cite{Ball:2012cx} and considering four additional PDF sets (CT14, MMHT2014, NNPDF3.0, and ATLAS-epWZ12), resulting in <1\% uncertainties in the signal acceptance. An uncertainty due to choosing the A14 tune for the signal generation is estimated by varying the scales for initial- and final-state radiation, as well as multiparton interactions, and results in an uncertainty of 2\% (3\%) for $WH$ ($ZH$) resonances.
 
Finally, an uncertainty in the Run 2 integrated luminosity of 1.7\%~\cite{ATLAS-CONF-2019-021} is considered, as obtained by the LUCID-2 detector~\cite{Avoni:2018iuv} for the main luminosity measurements. 
The impact of the main systematic uncertainties on signal event yields is summarized in Table~\ref{tab:syst}.
 
\begin{table}[t]
\caption{Systematic uncertainties affecting the signal event yields in the $WH$ and $ZH$
signal regions with 1-tag or 2-tag (denoted WH-1, WH-2, ZH-1, and ZH-2, respectively).
The HVT model is used with resonance masses of 2 and 4~\TeV.}
\begin{center}
\begin{tabular}{l|cccc|cccc}
\hline
\hline
Source & \multicolumn{4}{c|}{Signal (2~\TeV)} & \multicolumn{4}{c}{Signal (4~\TeV)} \\
& WH-1 & WH-2 & ZH-1 & ZH-2 & WH-1 & WH-2 & ZH-1 & ZH-2 \\ \hline
Jet energy scale        & 2.8\%~ & 3.1\%~ & 1.6\%~ & 2.1\%~ & 5.2\%~ & 8.4\%~ & 6.6\%~ & 8.0\%~ \\
Jet mass scale          & 18\% & 18\% & 9.6\%~ & 14\% & 20\% & 17\% & 21\% & 18\% \\
Jet mass resolution     & 43\% & 45\% & 37\% & 40\% & 22\% & 21\% & 21\% & 22\% \\
Flavor tagging          & 17\% & 8.4\%~ & 16\% & 8.4\%~ & 26\% & 12\% & 26\% & 13\% \\
$H$-boson tagging       & 10\% & 10\% & 10\% & 10\% & 10\% & 10\% & 10\% & 10\% \\
$V$-boson tagging       & 13\% & 13\% & 13\% & 13\% & 13\% & 13\% & 13\% & 13\% \\
Luminosity              & 1.7\%~ & 1.7\%~ & 1.7\%~ & 1.7\%~ & 1.7\%~ & 1.7\%~ & 1.7\%~ & 1.7\%~ \\
MC statistical uncertainty & 1.5\%~ & 1.4\%~ & 1.6\%~ & 1.4\%~ & 1.2\%~ & 1.4\%~ & 1.5\%~ & 1.7\%~ \\
\hline
\hline
\end{tabular}
\label{tab:syst}
\end{center}
\end{table}
 
% End of text imported from the .//sections/sec-systematics.tex input file

\section{Results}
\label{sec:result}
% The next lines are included from the .//sections/sec-results.tex input file
The statistical analysis of the data is performed using a binned likelihood function, constructed from the \mjj\ distributions in the 1-tag and 2-tag signal regions, using the procedure described in Ref.~\cite{HIGG-2012-27} and the \textsc{RooStats} framework~\cite{RooFit}. The \mjj\ histograms derived from MC simulation are used for the HVT $W^\prime$ and $Z^\prime$ processes, while the data-driven background estimates are used for the combined $t\bar{t}$, $V$+jets, and QCD multijet processes. The input \mjj\ distribution bounds are $[1.3, 6.0]$~\TeV.
 
A test statistic based on the profile likelihood ratio~\cite{Cowan:2010js} is used to test signal hypotheses, parameterized by the signal strength value, $\mu$, acting as a scale factor on the predicted number of signal events for each model assumption. The likelihood, $L$, is defined from the Poisson probability to observe $N$ data events for a given signal $s$ and background $b$ expectation in each bin of the final discriminant:
 
\begin{equation}\nonumber
L(\mu, \vec{\theta}) =
\prod_{c=1}^{\mathrm{categories}} \prod_{i=1}^{\mathrm{bins}}
\mathrm{Pois}\left(N_{ci} | \mu s_{ci}(\vec{\theta}) + b_{ci}(\vec{\theta})\right)
\prod_{j \in\,\vec{\theta}} f_{j}(\theta_{j}),
\end{equation}
where the index $c$ represents the 1-tag or 2-tag event categories and $i$ represents the bin in the \mjj\ distribution. Nuisance parameters $\vec{\theta}$ are included in the likelihood function with Gaussian or log-normal constraint terms, $f_{j}(\theta_{j})$. Those nuisance parameters which produce bin variations smaller than 1\% from the nominal value are neglected.
 
Experimental uncertainties in the signal are fully correlated between the 1-tag and 2-tag signal regions, whereas background modeling uncertainties are kept independent. In order to avoid an overconstraining of the background modeling uncertainties in the high mass region due to the higher statistical power at low masses, the \mjj\ shape uncertainties above and below 2.5~\TeV\ are allowed to vary independently in the fit. The post-fit background expectation was found to be stable and independent of the particular choice of splitting point.
 
The fits are performed separately for the $W^\prime$ and $Z^\prime$ models, using data in the 1-tag and 2-tag regions from the SRWH and SRZH selections, respectively. The fit results are interpreted independently for the $W^\prime$ and $Z^\prime$ hypotheses~--~the $WH$ and $ZH$ signal regions are not orthogonal and have approximately 40\% of events in common, in each category.
 
The pre- and post-fit \mjj\ distributions in the signal region are shown in Figure~\ref{fig:mVH_SR}, for signal resonances with a mass of 2~\TeV. The numbers of data events in the signal regions are shown in Table~\ref{tab:yields_SR}, along with the predicted background and signal yields, post-fit. No events with \mjj\ values above 5~TeV are selected.
 
\begin{figure}[tbp!]
\begin{center}
\includegraphics[width=0.48\textwidth]{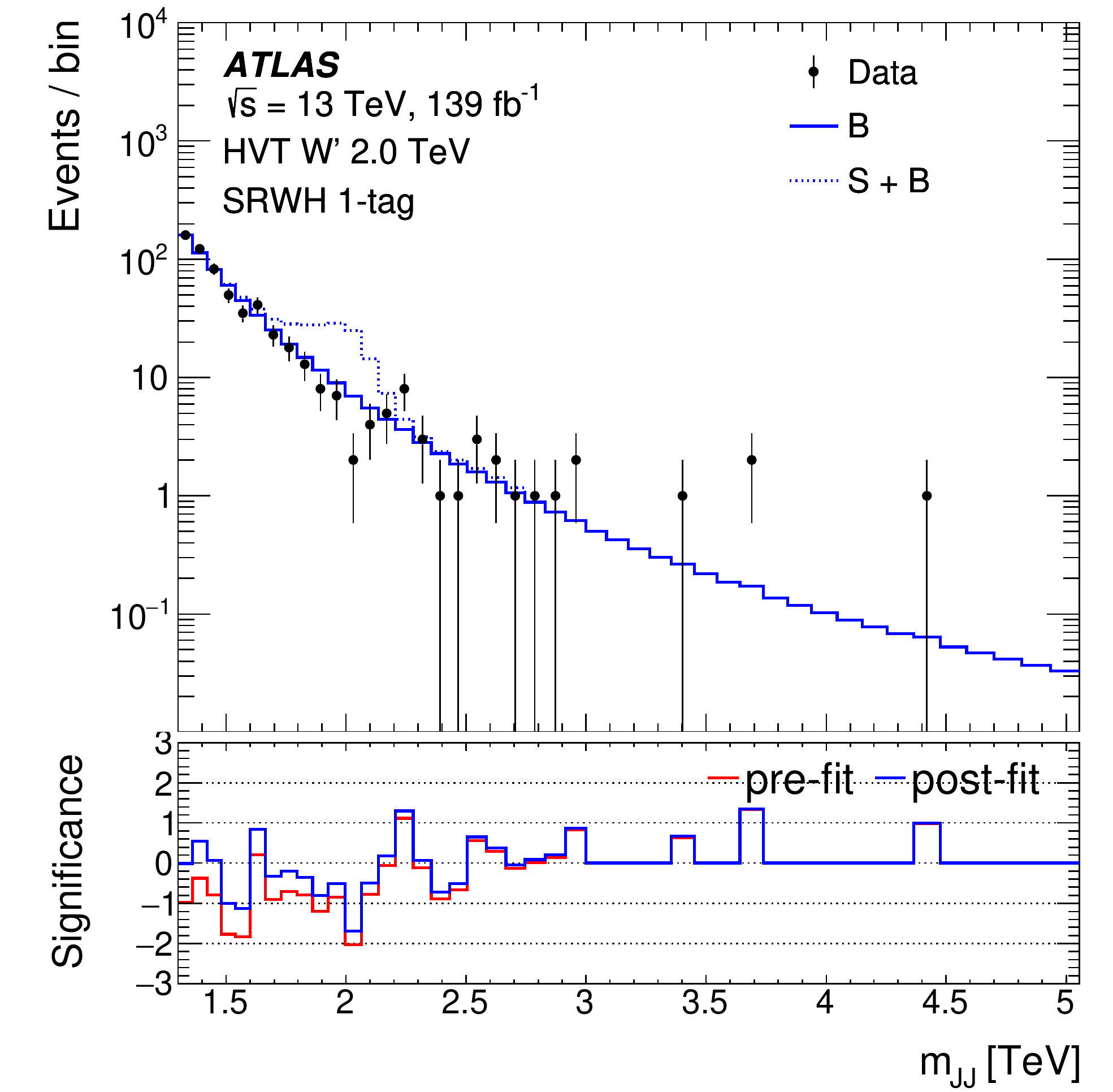}
\includegraphics[width=0.48\textwidth]{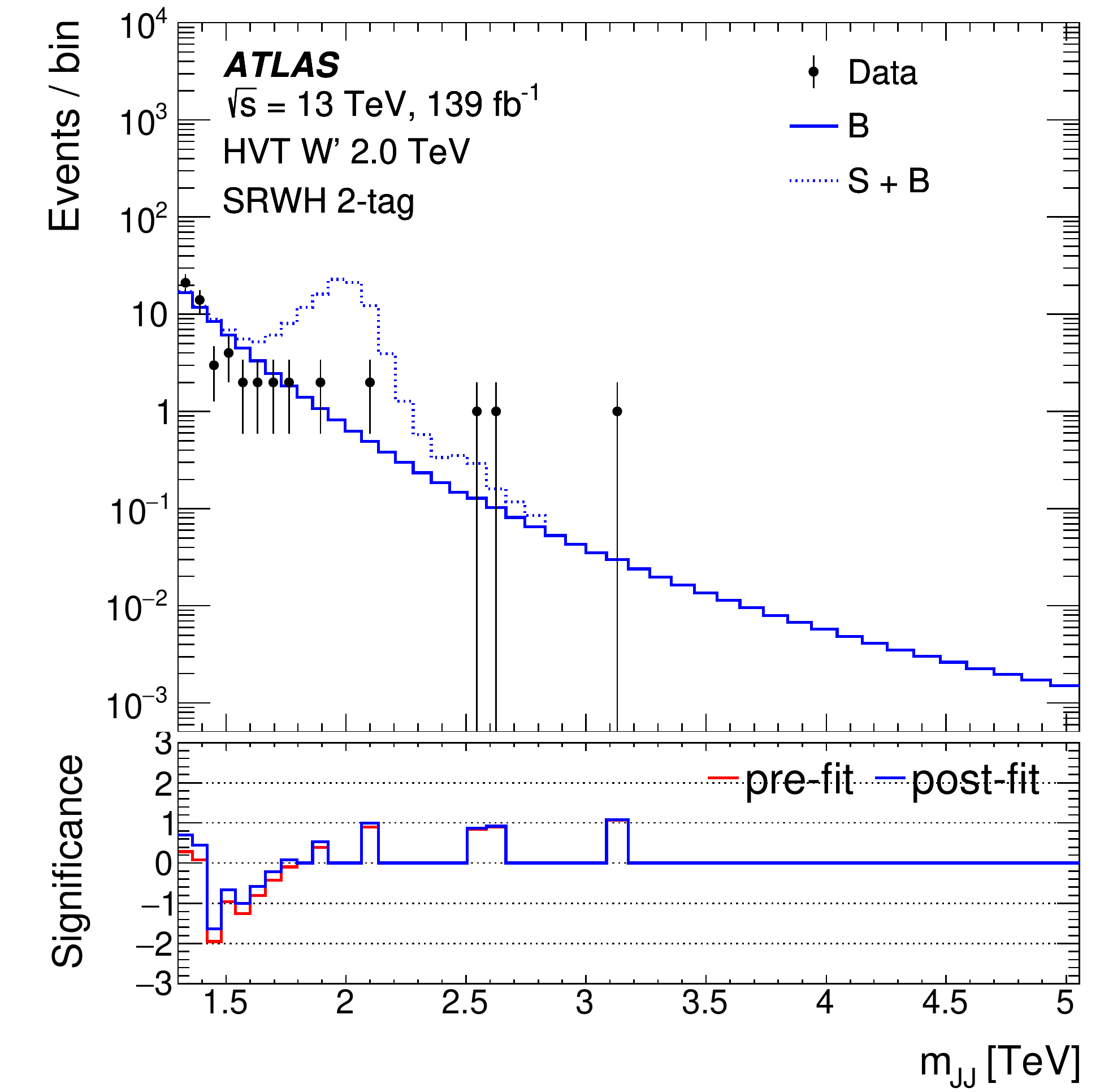}
\includegraphics[width=0.48\textwidth]{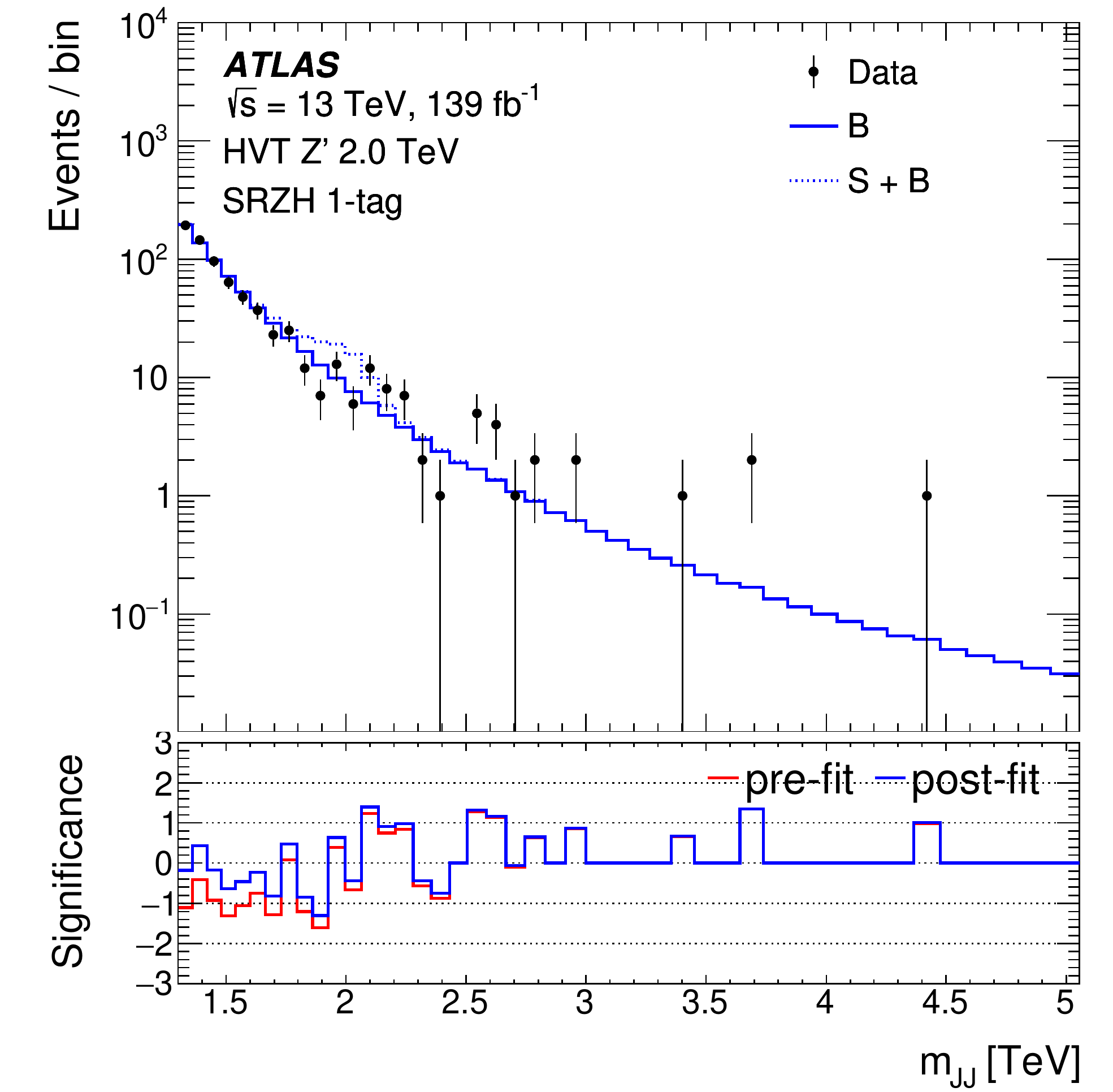}
\includegraphics[width=0.48\textwidth]{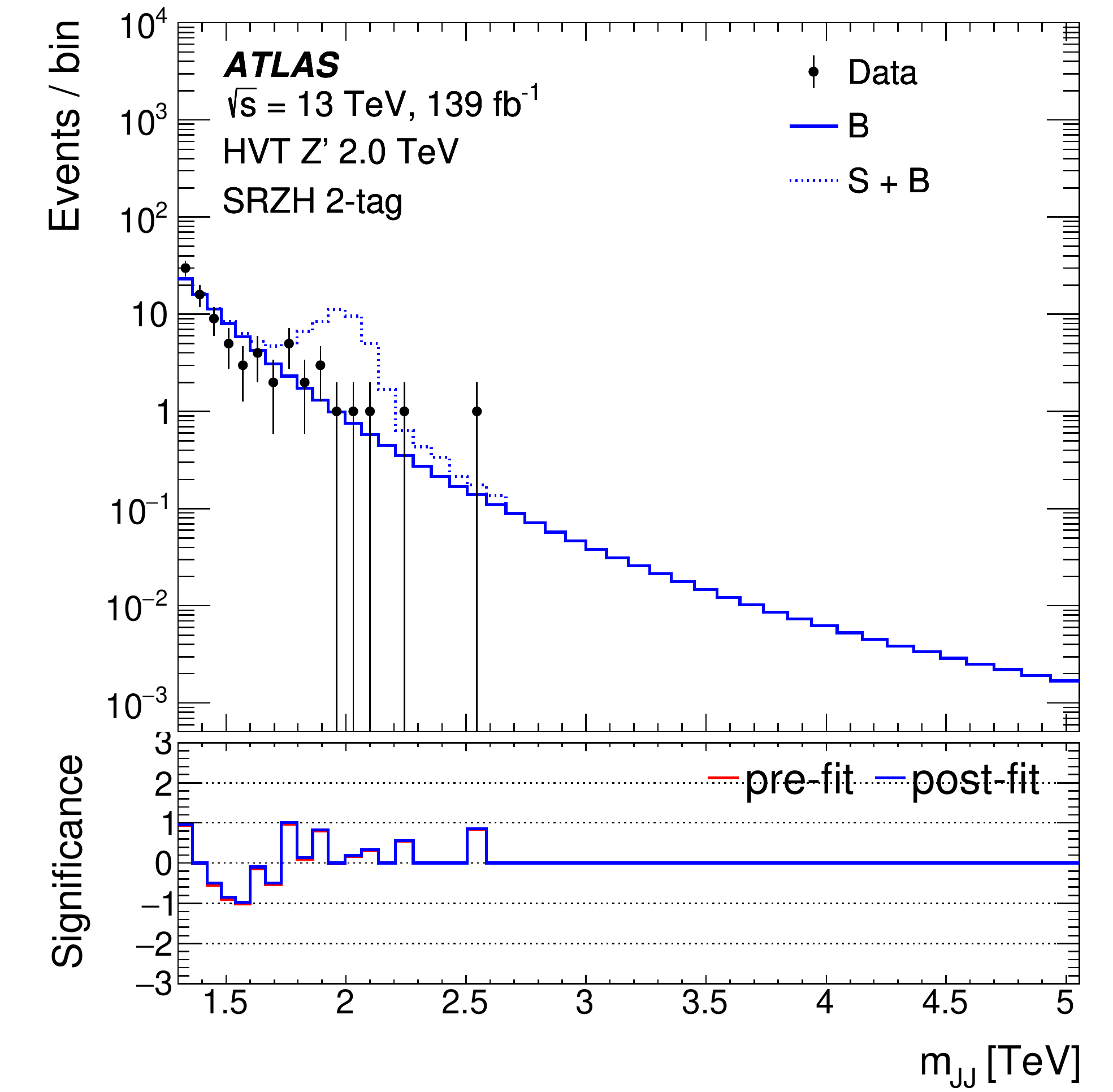}
\end{center}
\caption{Dijet mass distributions in the $WH$ (top) and $ZH$ (bottom) signal regions, after the likelihood fit to events in the 1-tag (left) and 2-tag (right) categories. The black points correspond to data and the solid blue histogram to the post-fit background prediction. The $WH$ and $ZH$ signal regions are not orthogonal.
The expected signal distributions for a $V^\prime$ boson with mass of 2~\TeV\ are also shown (dashed histograms). The bin width varies and corresponds to the experimental mass resolution. Distributions of the significance of the observed deviations from the expected background are presented in the bottom panels before and after the fit. The significance calculation assumes Poisson probabilities and only accounts for statistical fluctuations.}
\label{fig:mVH_SR}
\end{figure}

\begin{table}[t]
\caption{Event yields for the data, predicted background, and signal in each of the signal regions. The signal corresponds to that expected for HVT model B with resonance masses of 2 and 4~\TeV.}
\begin{center}
\begin{tabular}{ccccc}
\hline
\hline
Region & Data & Background & Signal (2~\TeV) & Signal (4~\TeV) \\ \hline
SRWH 1-tag & 598 & $612 \pm 46$ & 110~~ & 2.0~~ \\
SRWH 2-tag &  ~~57 & ~~$61 \pm 13$ & 100~~ & 1.0~~ \\
SRZH 1-tag~ & 717 & $725 \pm 53$ &  47 & 0.80 \\
SRZH 2-tag~ &  ~~84 & ~~$81 \pm 17$ &  44 & 0.42 \\
\hline
\hline
\end{tabular}
\label{tab:yields_SR}
\end{center}
\end{table}
 
A test of the background-only hypothesis is performed by setting $\mu$ equal to zero in the likelihood fit. Deviations from the background-only hypothesis are quantified by determining the local $p_{0}$-value at each signal mass point. The largest deviation is observed in the fit to the $WH$ signal regions and corresponds to a $p_{0}$-value of 0.03 for a resonance mass of 2.8~\TeV.
 
The data are used to set upper limits on the production cross section of new resonances. Exclusion limits are computed using the CL$_\text{s}$ method~\cite{Read:2002hq}, with a value of $\mu$ regarded as excluded at the 95\% CL when the CL$_\text{s}$ value is less than 5\%. The cross-section limits are shown in Figure~\ref{fig:limits}. The observed limits range from cross sections of \maxwhxs~fb to \minwhxs~fb for $WH$ and from \maxzhxs~fb to \minzhxs~fb for $ZH$, corresponding to the exclusion of $W^\prime$ ($Z^\prime$) resonances up to a mass of \wpmassA~\TeV\ (\zpmassA~\TeV) in the context of HVT model A and \wpmassB~\TeV\ (\zpmassB~\TeV) in the context of HVT model B. The 2-tag category dominates the sensitivity of the search at low resonance mass while the impact of the 1-tag category increases at higher mass, surpassing the 2-tag category at masses above 3.9 TeV. As a test of the asymptotic approximation used in the statistical analysis, results are also obtained with ensembles of pseudoexperiments. The cross-section upper limits obtained in that case are looser by 10--20\%, with a larger difference at high $m(V^\prime)$, and the mass limits are at most 0.05~\TeV\ weaker.
 
These results can also be translated into exclusions in the \{$g_{H}, g_{f}$\} plane, where $g_f$ represents a universal coupling between the $V^\prime$ bosons and fermions. Here, $g_q$ is taken to be equal to $g_f$. Figure~\ref{fig:couplings} shows the 95\% CL limits in this plane for several resonance masses.
 
\begin{figure}[tbp!]
\begin{center}
\includegraphics[width=0.48\textwidth]{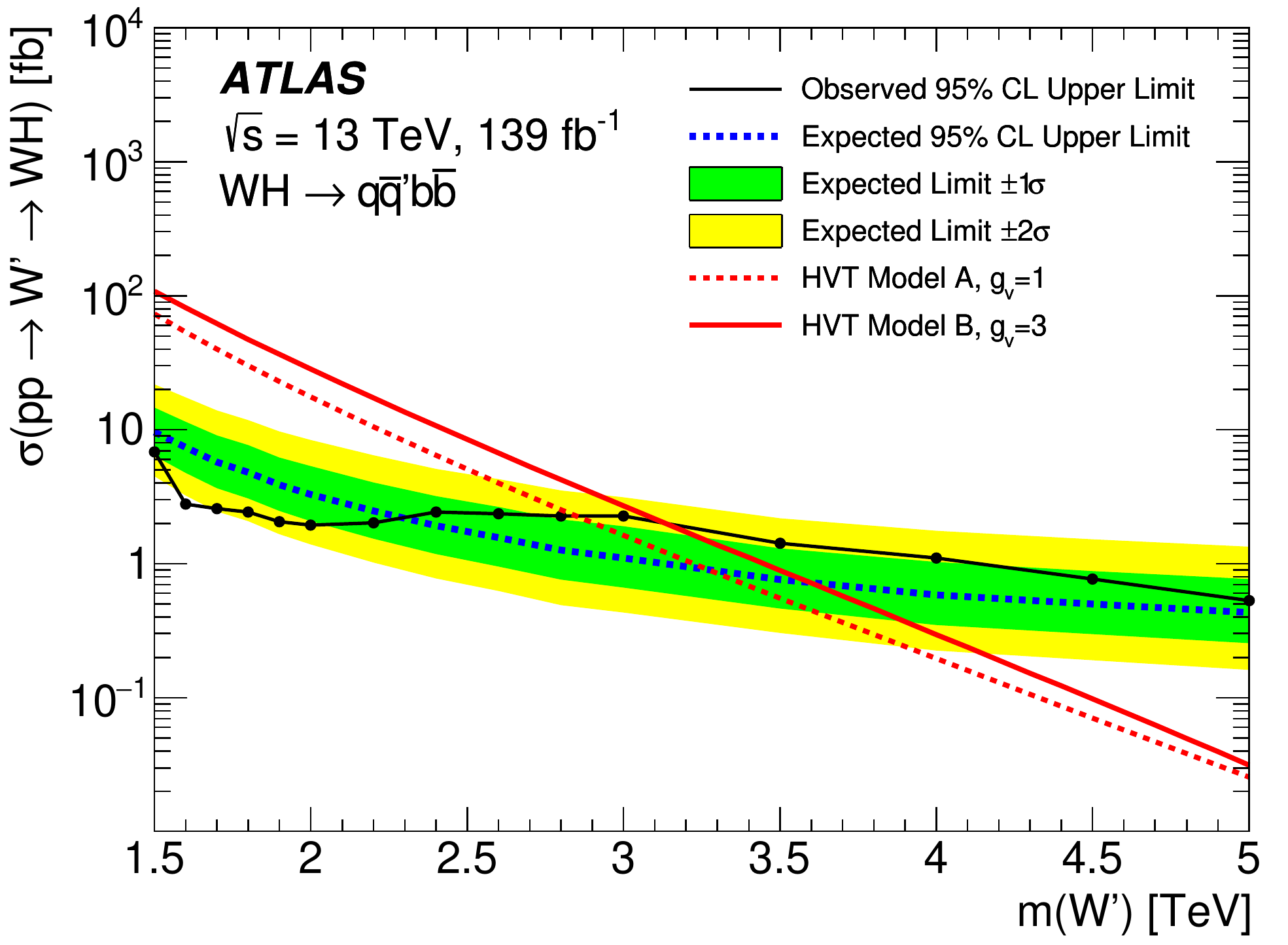}
\includegraphics[width=0.48\textwidth]{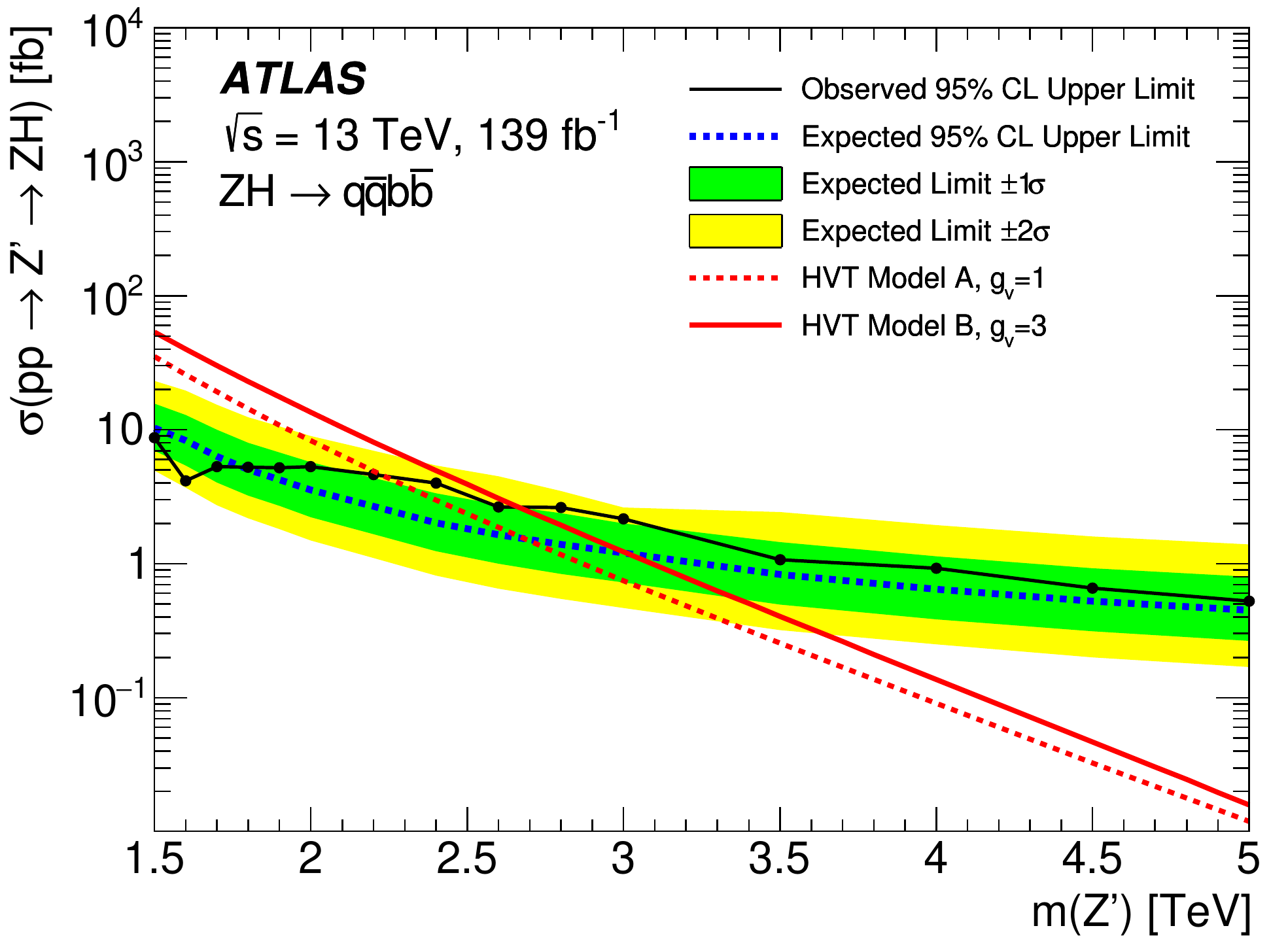}
\end{center}
\caption{Observed and expected 95\% CL upper limits on the cross section for $pp\rightarrow V^\prime \rightarrow VH$ 
in the $WH$ (left) and $ZH$ (right) channels. The red solid (dashed) lines show the cross-section predictions as a function of the resonance mass in the context of HVT model B (A). }
\label{fig:limits}
\end{figure}
 
\begin{figure}[htbp!]
\begin{center}
\includegraphics[width=0.48\textwidth]{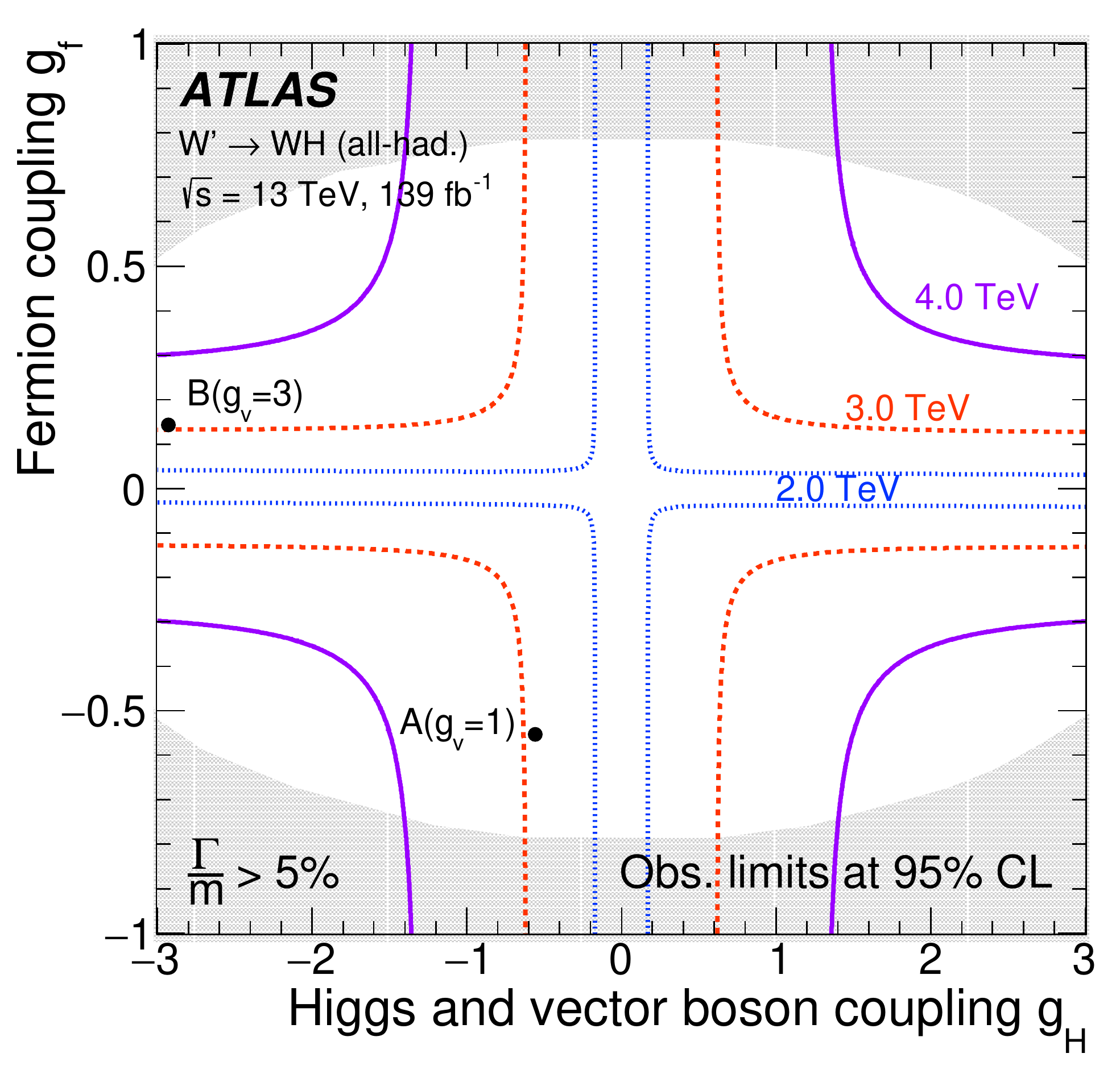}
\includegraphics[width=0.48\textwidth]{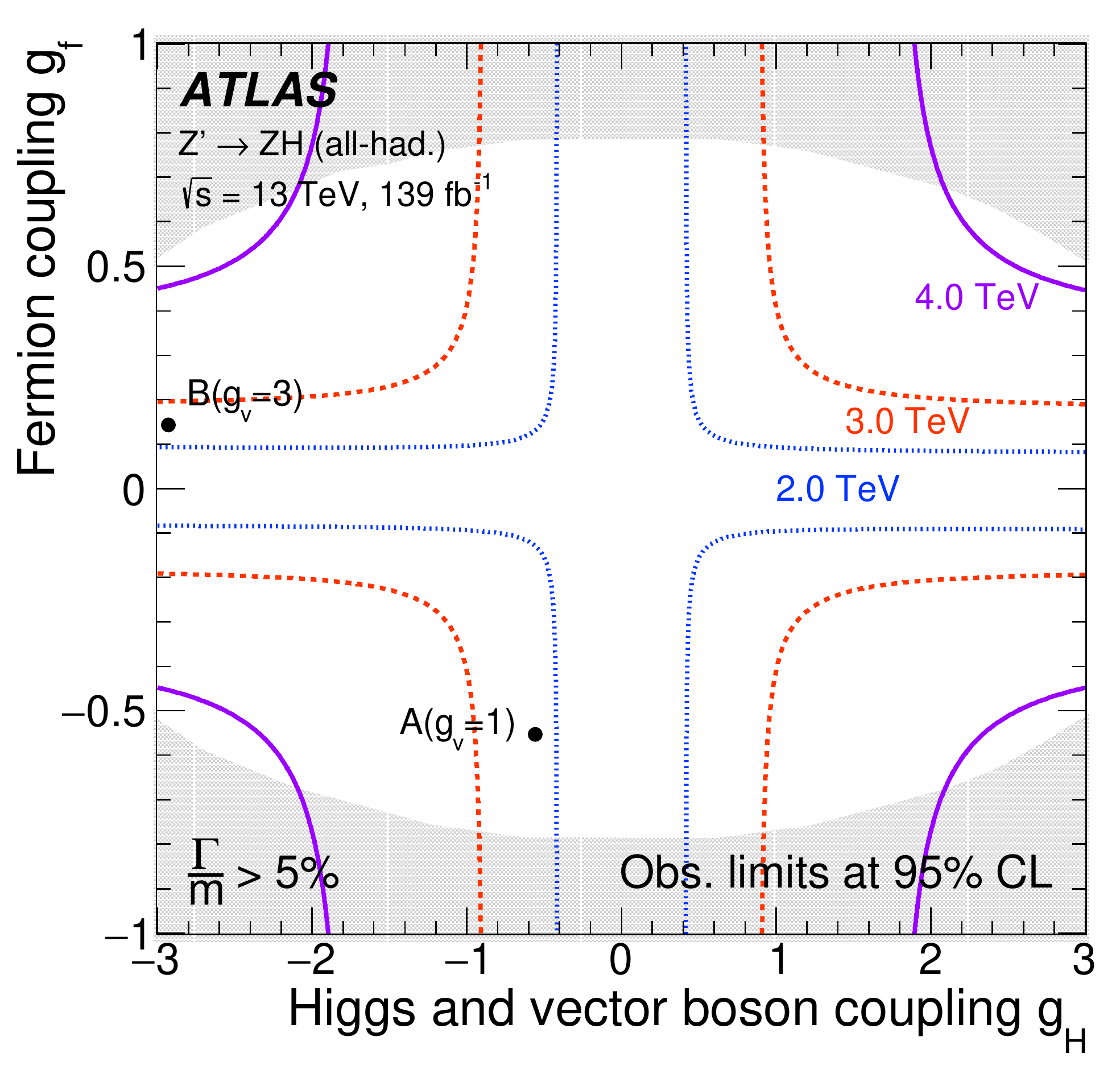}
\end{center}
\caption{Limits at 95\% CL in the $g_{f}$ vs.\ $g_{H}$ plane for resonance masses of 2, 3, and 4~\TeV\ for the $WH$ (left) and $ZH$ (right) channels in the context of the HVT model. The coupling values corresponding to HVT models A and B are indicated by filled circles. The gray region indicates values of the couplings corresponding to $V^{\prime}$ resonances with $\Gamma/m$ greater than 5\%. In that region, the assumption that the $V^{\prime}$ signal \mjj\ shape is dominated by the experimental resolution is no longer valid.}
\label{fig:couplings}
\end{figure}
% End of text imported from the .//sections/sec-results.tex input file
 
\FloatBarrier
 
\section{Conclusion}
% The next lines are included from the .//sections/sec-conclusion.tex input file
 
A search for heavy resonances decaying into a $W$ or $Z$ boson and a Higgs boson
is reported. The results are based on a sample of $pp$ collisions at $\rts = 13$~\TeV\
collected by the ATLAS detector at the Large Hadron Collider, corresponding to 139~\ifb\ of integrated luminosity.
The search exploits jet substructure techniques to study the fully hadronic $qqbb$
final state which results from the dominant decay modes of the $W/Z$ and Higgs bosons.
The main background contribution arises from multijet production, with a smaller
contribution from top-quark pair and $V$+jets production. All background contributions are extracted directly
from the data. Compared with the previously available results, the search benefits from an increased integrated luminosity, as well as from improvements in reconstruction and tagging of large-$R$ jets and track-jets with \pt-dependent radius.
 
No significant excess of events is observed over the expected background and
the upper limits set on the cross section for $pp \to W^\prime \to W H$
and $pp \to Z^\prime \to Z H$ 
range from \maxwhxs~fb at $m(W^\prime) = 1.5$~\TeV\ to \minwhxs~fb at $m(W^\prime) = 5.0$~\TeV,
and from \maxzhxs~fb at $m(Z^\prime) = 1.5$~\TeV\ to \minzhxs~fb at $m(Z^\prime) = 5.0$~\TeV,
at 95\% CL.
These results translate into lower limits on the mass of $W^\prime$ ($Z^\prime$) bosons
of \wpmassA~\TeV\ (\zpmassA~\TeV) in the context of the weakly coupled HVT model A and
of \wpmassB~\TeV\ (\zpmassB~\TeV) in the context of the strongly coupled HVT model B, at 95\% CL.

% End of text imported from the .//sections/sec-conclusion.tex input file

\section*{Acknowledgments}
 
% The next lines are included from the .//acknowledgements/Acknowledgements.tex input file

We thank CERN for the very successful operation of the LHC, as well as the
support staff from our institutions without whom ATLAS could not be
operated efficiently.
 
We acknowledge the support of ANPCyT, Argentina; YerPhI, Armenia; ARC, Australia; BMWFW and FWF, Austria; ANAS, Azerbaijan; SSTC, Belarus; CNPq and FAPESP, Brazil; NSERC, NRC and CFI, Canada; CERN; ANID, Chile; CAS, MOST and NSFC, China; COLCIENCIAS, Colombia; MSMT CR, MPO CR and VSC CR, Czech Republic; DNRF and DNSRC, Denmark; IN2P3-CNRS and CEA-DRF/IRFU, France; SRNSFG, Georgia; BMBF, HGF and MPG, Germany; GSRT, Greece; RGC and Hong Kong SAR, China; ISF and Benoziyo Center, Israel; INFN, Italy; MEXT and JSPS, Japan; CNRST, Morocco; NWO, Netherlands; RCN, Norway; MNiSW and NCN, Poland; FCT, Portugal; MNE/IFA, Romania; MES of Russia and NRC KI, Russia Federation; JINR; MESTD, Serbia; MSSR, Slovakia; ARRS and MIZ\v{S}, Slovenia; DST/NRF, South Africa; MICINN, Spain; SRC and Wallenberg Foundation, Sweden; SERI, SNSF and Cantons of Bern and Geneva, Switzerland; MOST, Taiwan; TAEK, Turkey; STFC, United Kingdom; DOE and NSF, United States of America. In addition, individual groups and members have received support from BCKDF, CANARIE, Compute Canada and CRC, Canada; ERC, ERDF, Horizon 2020, Marie Sk{\l}odowska-Curie Actions and COST, European Union; Investissements d'Avenir Labex, Investissements d'Avenir Idex and ANR, France; DFG and AvH Foundation, Germany; Herakleitos, Thales and Aristeia programmes co-financed by EU-ESF and the Greek NSRF, Greece; BSF-NSF and GIF, Israel; La Caixa Banking Foundation, CERCA Programme Generalitat de Catalunya and PROMETEO and GenT Programmes Generalitat Valenciana, Spain; G\"{o}ran Gustafssons Stiftelse, Sweden; The Royal Society and Leverhulme Trust, United Kingdom.
 
The crucial computing support from all WLCG partners is acknowledged gratefully, in particular from CERN, the ATLAS Tier-1 facilities at TRIUMF (Canada), NDGF (Denmark, Norway, Sweden), CC-IN2P3 (France), KIT/GridKA (Germany), INFN-CNAF (Italy), NL-T1 (Netherlands), PIC (Spain), ASGC (Taiwan), RAL (UK) and BNL (USA), the Tier-2 facilities worldwide and large non-WLCG resource providers. Major contributors of computing resources are listed in Ref.~\cite{ATL-SOFT-PUB-2020-001}.
 
% End of text imported from the .//acknowledgements/Acknowledgements.tex input file

\clearpage

\printbibliography
\clearpage % ATLAS Collaboration author list
% Reference date of HDBS-2018-11 is 2020-04-22
% Author list last updated on date 04-JAN-21
% Data extracted on 04-Jan-2021 for paper reference HDBS-2018-11
% at 8:48pm
 
\begin{flushleft}
\hypersetup{urlcolor=black}
{\Large The ATLAS Collaboration}

\bigskip

\AtlasOrcid[0000-0002-6665-4934]{G.~Aad}$^\textrm{\scriptsize 102}$,    
\AtlasOrcid[0000-0002-5888-2734]{B.~Abbott}$^\textrm{\scriptsize 128}$,    
\AtlasOrcid{D.C.~Abbott}$^\textrm{\scriptsize 103}$,    
\AtlasOrcid[0000-0002-2788-3822]{A.~Abed~Abud}$^\textrm{\scriptsize 36}$,    
\AtlasOrcid[0000-0002-1002-1652]{K.~Abeling}$^\textrm{\scriptsize 53}$,    
\AtlasOrcid[0000-0002-2987-4006]{D.K.~Abhayasinghe}$^\textrm{\scriptsize 94}$,    
\AtlasOrcid[0000-0002-8496-9294]{S.H.~Abidi}$^\textrm{\scriptsize 167}$,    
\AtlasOrcid[0000-0002-8279-9324]{O.S.~AbouZeid}$^\textrm{\scriptsize 40}$,    
\AtlasOrcid{N.L.~Abraham}$^\textrm{\scriptsize 156}$,    
\AtlasOrcid[0000-0001-5329-6640]{H.~Abramowicz}$^\textrm{\scriptsize 161}$
% ...
\bigskip
\\

% ...
$^{36}$CERN, Geneva; Switzerland.\\
% ...
$^{40}$Niels Bohr Institute, University of Copenhagen, Copenhagen; Denmark.\\
% ...
$^{53}$II. Physikalisches Institut, Georg-August-Universit\"{a}t G\"ottingen, G\"ottingen; Germany.\\
% ...
$^{94}$Department of Physics, Royal Holloway University of London, Egham; United Kingdom.\\
% ...
$^{102}$CPPM, Aix-Marseille Universit\'e, CNRS/IN2P3, Marseille; France.\\
$^{103}$Department of Physics, University of Massachusetts, Amherst MA; United States of America.\\
% ...
$^{128}$Homer L. Dodge Department of Physics and Astronomy, University of Oklahoma, Norman OK; United States of America.\\
% ...
$^{156}$Department of Physics and Astronomy, University of Sussex, Brighton; United Kingdom.\\
% ...
$^{161}$Raymond and Beverly Sackler School of Physics and Astronomy, Tel Aviv University, Tel Aviv; Israel.\\
% ...
$^{167}$Department of Physics, University of Toronto, Toronto ON; Canada.\\
% ...

\end{flushleft}

% Created with Glance <Atlas.Glance@cern.ch>

\end{document}